# How to interpret algorithmically constructed topical structures of research specialties? A case study comparing an *internal* and an *external* mapping of the topical structure of invasion biology

Matthias Held[1] and Theresa Velden[2]

[1] *held@ztg.tu-berlin.de*
Zentrum für Technik und Gesellschaft, Technische Universität Berlin, Hardenbergstr. 16-18, D-10623 Berlin (Germany)

[2] *velden@dzhw.eu*
Deutsches Zentrum für Hochschul- und Wissenschaftsforschung Berlin, Schützenstraße 6a, 10117 Berlin (Germany)

**Abstract**
In our paper we seek to address a shortcoming in the scientometric literature, namely that, given the proliferation of algorithmic approaches to topic detection from bibliometric data, there is a relative lack of studies that validate and create a deeper understanding of the topical structures these algorithmic approaches generate. To take a closer look at this issue, we investigate the results of the new Leiden algorithm when applied to the direct citation network of a field-level data set. We compare this 'internal' perspective which is constructed from the citation links within a data set of 30,000 publications in invasion biology, with an 'external' perspective onto the topic structures in this research specialty, which is based on a global science map in form of the CWTS microfield classification underlying the Leiden Ranking. We present an initial comparative analysis of the results and lay out our next steps that will involve engaging with domain experts to examine how the algorithmically identified topics relate to understandings of topics and topical perspectives that operate within this research specialty.

**Introduction**
While algorithms and their application to extract topical structures from bibliometric data proliferate, there is a shortage of studies that validate their results and contribute to a deeper understanding of the variation in topical structures that these algorithmic approaches create. So far only a small set of studies exists that systematically investigate the validity of solutions obtained and the difference made by alternative choices (see e.g. Haunschild et al. 2018, Sjögårde 2018, Klavans & Boyack 2017, Velden et al. 2017, Šubelj et al 2016, Boyack & Klavans 2010, Klavans & Boyack 2011, Shibata et al. 2009). We are concerned that failure to invest into the systematic comparisons and careful validation of the interpretation of algorithmically extracted topical structures undermines our ability to provide robust interpretations of their results and sound guidance on the choice and appropriate use of algorithmic topic extraction or field classification approaches. This study is a contribution to what we conceive of as a larger, much needed, research program. This program should address, on the one hand, the question of the constructive nature of a topic extraction result, relying on decisions on dataset, data model, algorithm and its parameters (Gläser et al. 2017). On the other hand, it should address the validity of interpretations of results by examining the degree of agreement between a theoretical definition of the topic concept (Havemann et al. 2017) with the actual operationalization through the chosen approach, and by exploring their correspondence to perceptions of topical structures held by their creators "in-the-wild", the researchers themselves.

This paper is a case study that takes a closer look at the topical structures obtained when using the newly released Leiden algorithm for community detection (Traag et al. 2018) to produce a local map of the field of invasion biology. As a first step, we compare this 'internal' perspective that is based exclusively on relations in the direct citation network[1] of approx. 30,000 publications in invasion biology, with an 'external' perspective that is generated by projecting this data set of publications onto a global map of science. We use the global map that underlies the field classification of the Leiden Ranking[2] and consists of approx. 20 mio publications grouped into 4047 microfields. It has been produced by CWTS using the Smart Local Moving algorithm for community detection (Waltman & Van Eck 2013). Such an external perspective captures the embedding of publications in a field into the global network of scientific publications and is expected to highlight interdisciplinary connections to other areas of research (Boyack 2017).

Klavans and Boyack (2011) argue that under certain conditions[3] a global science map can be expected to produce a more 'accurate' map of a field than local maps can - where accuracy is measured by textual coherence of the clusters obtained. However, as Haunschild et al. (2018) found in a case study of the topic of 'overall water splitting', also global maps may fail to adequately capture research fields. In this study, given the sparseness of evidence so far, rather than dismiss the local map as less accurate per se, we keep an open mind. Our interest is to investigate the capability of either perspective, internal or external, to capture understandings of topics that operate within the research specialty of invasion biology[4]. In the following we present an initial bibliometric comparison of the results obtained with these two mapping approaches, and discuss our next steps that will involve engaging with field experts who do research in invasion biology to discuss interpretations of the results of these alternative algorithmic mappings of their field of research.

**Data & Methods**

In this study we use a data set that is based on a lexical query developed by researchers in invasion biology (Vaz et al. 2017) in order to capture publications belonging to their research specialty:

"Ecological invasion*" or "Biological invasion*" or "Invasion biology" or "Invasion ecology" or "Invasive species" or "Alien species" or "Introduced species" or "Non-native species" or "Nonnative species" or "Nonindigenous species" or "Non-indigenous species" or "Allochthonous species" or "Exotic species".

Using the lexical query above, 30,731 document IDs from the Web of Science database were retrieved on August 28, 2017 (Figure 1). For this set, we were able to retrieve the relevant metadata (titles, abstracts, source, publication year, document types, cited references) from the 2018 stable version of the Web of Science database hosted by the 'Kompetenzzentrum Bibliometrie' (KB). We decided to restrict the time window to the years 2000-2017[5], and the document types to article, letters and reviews. The further analysis was done using the giant component of the direct citation network. The network of the remaining 25,680 publications and 229,572 citation links[6] served as input for the Leiden algorithm and the projection onto the CWTS microfield classification (explained below).

---

[1] Direct citation networks are a popular choice in bibliometrics for the production of global science maps, given their relative sparseness. Previous studies (Klavans & Boyack 2017, Velden et al. 2017a, Shibata et al. 2009) suggest its usefulness to extract taxonomic topic structures.

[2] http://www.leidenranking.com/

[3] "All things being equal", meaning if data source, data model, algorithm and so forth are the same. They also suggest that local maps will be less accurate in particular if boundary forces (links to concepts outside the field) are stronger than core forces.

[4] Our interest in validating algorithmically generated science maps derives from theory-guided empirical work we are engaged in: M. Held is involved in the project MIMAL that explores linkages between bibliometric patterns at the micro level and the macro level; T. Velden is involved in the project 'Field specific forms of open science' that compares four fields of science and uses bibliometric maps of research specialties to support comparisons in ethnographic science studies.

[5] A specific reason for this choice was the desire to increase comparability with another bibliometric study of the field by Enders et al, (in preparation), as well as the general consideration that the field of invasion biology has experienced critical growth in the early 2000's such that the large majority of publications in the field is included in the chosen time window.

For clustering the direct citation network, we chose the recently released Leiden algorithm (Traag et al. 2018), a community detection algorithm which has been developed to overcome a decisive shortcoming of a widely used community detection algorithm, the Louvain algorithm (Blondel et al. 2008), namely the production of badly connected clusters. It further avoids the use of modularity as quality function due to its known shortcoming of a resolution limit and instead chooses the quality function Constant Potts Model (CPM) that has been shown to be resolution-limit free (Traag et al. 2011, 2018). For the CPM we chose two resolution values and minimum cluster sizes. Different from the methodology introduced in Waltman & Van Eck (2012), we do not merge clusters below the threshold, and instead discard them. The publications from those discarded clusters amount to less than 10% of the publications in both solutions Leiden$_6$ and Leiden$_{16}$. The algorithm was started with a random seed, run with 100 iterations with ten random starts each.

**Figure 1: Schematic representation of the processing steps.**

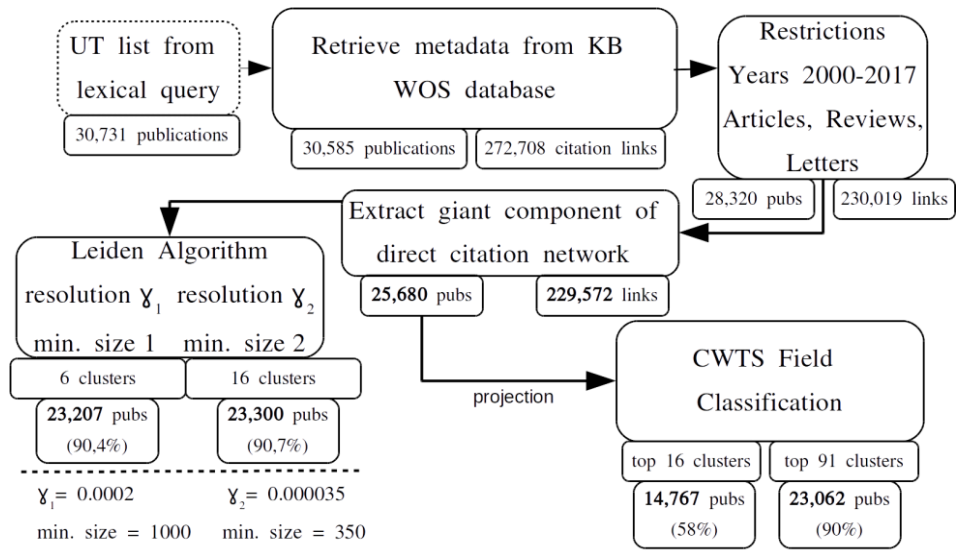

To contrast this 'internal' perspective of a clustering of a research specialty with an 'external' perspective that takes the embedding of publications in the global citation network of science into account, we project our field data set onto the CWTS microfield classification. It consist of 4047 microfields that have been extracted with the SLM algorithm on the weighted direct citation network of more than 20 million publications published in 2000-2017 and indexed by the Web of Science[7]. Of the 25,680 UTs included in the giant component, 25,627 can be found in the micro fields of the CWTS field classification. We defined as clusters in our projection cluster solution the largest intersections between our field data set and a microfield (in terms of absolute number of publications).

Finally, in order to find characteristic terms to describe the content of clusters, we extracted the noun phrases from titles and abstracts of the publications of each cluster in each cluster solution, using part of speech tagging and chunking available in the Python package 'nltk'. Terms which had been used in the lexical query to delineate the field were excluded. To obtain a measure for how well each of the remaining terms describes the content of each cluster, we used the differential cluster labelling by Koopman & Wang (2017), which is based on normalized mutual information (NMI). The higher the value, the more significant the term to characterize the cluster and differentiate it from the rest. For labelling the clusters in the Leiden$_6$ and Leiden$_{16}$ solution, the terms of all publications from the giant component were included. In order to label the clusters in the projection solution, we only included terms and publications which occurred in the projection clusters. An additional set of labels was produced with the same approach, using journal names instead of extracted terms. The cluster labels eventually used in this paper were manually derived from those NMI score ranked lists of terms and journals by considering information provided on habitat, organisms, research problem, and the subject area of journals.

---

[6] Following Waltman & Van Eck (2012), we produced a weighted version of the direct citation network to account for a potential variation of in citation practices within the field of invasion biology.

[7] http://www.leidenranking.com/information/fields (Accessed January 25, 2019)

In order to visualize the relationships between the topics identified, we use topic affinity networks that evaluate the strength of citation links between clusters to determine the affinity between topics. The existence of a link between topics in the affinity network indicates a surplus of connectivity between the two topics compared to a random null model (see Velden, Yan & Lagoze 2017 for details).

**Results**

Topics and sizes of the 6, respectively 16 clusters in the Leiden$_6$ /Leiden$_{16}$ solutions are given in the data appendix of this paper. An analysis of how the publications are regrouped from the six clusters of the Leiden$_6$ solution to the 16 clusters of the Leiden$_{16}$ solution suggests a continuity of the overall topic structure extracted by the two clustering runs. While two topics, 'marine invasion' (Leiden$_6$ C3) and 'trees and pests' (Leiden$_6$ C5) are largely preserved, other clusters get split into smaller, refined topics. Given that the two solutions are independent and not the result of a hierarchical clustering approach, this seems noteworthy and encouraging regarding an internal consistency of results achieved with the Leiden algorithm at different levels of resolution.

Sizes and topics of some of the larger projection clusters, as well as information of the CWTS microfields that the projection clusters are embedded in are also given in the data appendix. But for a few exceptions, the projection clusters constitute only about 5% of a microfield in the CWTS classification. Adopting the terminology used by Klavans and Boyack (2011) when comparing a local and a global mapping of the field of information science, microfield m402 may be considered a 'core' microfield for invasion science (53% of its publications overlap with the invasion data set and constitute projection cluster C1 on 'invasive plants'). Three microfields may be considered 'boundary' microfields, namely m2749 (34% overlap, projection cluster C5 on 'marine aquatic invasion, ballast water, ascidians'), m1774 (17% overlap, projection cluster C2 on 'freshwater aquatic invasion, great lakes'), and m2568 (17% overlap, projection cluster C10 on 'freshwater aquatic invasion, crayfish'). All other microfields may be considered 'boundary-crossing', i.e. largely outside the field of investigation.

In Figure 2 we compare the Leiden$_{16}$ with the projection cluster solution based on their topic affinity networks. Both solutions agree in that they include a cluster related to 'invasive plants' that consists of almost 25% of publications in the giant component. They differ in that the cluster size concentration of the projection cluster solution is lower: to cover a similar large proportion of publications from the giant component as the Leiden$_{16}$ solution (> 90%), one has to include the 91 largest clusters of the projection cluster solution, down to a size of 37 publications. The alluvial diagram in Figure 3 shows the regrouping of publications between the Leiden$_{16}$ solution and the 91 largest topics in the projection solution. While some topic continuity can be observed and the core of some topics clearly persists, other topics get fragmented. The zoom-in in Figure 3 shows how the topics of 'marine aquatic invasion' (C2 Leiden$_{16}$) and 'ballast water' (C15 Leiden$_{16}$) split-up into numerous topics in the projection solution (these topics are located next to each other in the area of the affinity network of the projection solution circled in Figure 2). We offer two observations: first, a technical one, namely that our labelling procedure fails to extract hints of organisms or research angle when cluster sizes fall below about 100 documents (see labels in Fig. 3 for C54, C56 or C58). Second, the refinement of topics often seems driven by a focus on a specific organism: crab, algae, oyster, jellyfish, etc. Occasionally, it is driven by a specific habitat: Mediterranean, Suez canal, Antarctica.

**Discussion & Future Work**

The alternative internal and external mappings in this paper provide two perspectives onto the topical structure of the field of invasion biology. Common feature of those topical structures is the concentration of almost 25% of publications in a topic relating to invasive plants. This is in alignment with statements by experts that suggest that observational studies of terrestrial plants dominate the empirical literature in the field (Jeschke & Heger 2018, p. 162). Further, organisms and habitat seem to constitute important dimensions for delineating topics - likely not a surprising observation for a domain expert. Notable exceptions are the projection based topics on 'genetic diversity' (C12), and 'models, climate change' (C7) that seem rather method or research angle-driven.

The projection of the invasion biology data set onto microfields of the global map promises insights into links between topics in invasion science to neighbouring fields. However, it suffers from an enormous spread. One third of publications are represented by 'invasive plants' as a core topic and three aquatic invasion-related boundary topics. The remaining ⅔ of publications are associated with and embedded in hundreds of different microfields with each less than a 10% share. This spread of invasion biology publications across microfields suggests that a delineation and representation of the research field of invasion biology through selection of a microfield from the global map might be ill-conceived - given one trusts the lexical query used by us and by Vaz et al. (2017) to delineate the field of invasion biology. This aligns with the finding by Haunschild et al. (2018) on the field of overall water splitting. The question what these microfields are representative of and their role as a topical context for research in a research specialty such as invasion biology, deserves further investigation.

Before moving on, we plan to improve our labelling approach by implementing entity recognition of taxonomic species. This way we expect to increase the meaningfulness and precision of the content labels we extract, allowing us e.g. to contrast the content of the projection clusters with the other publications in the microfield they are embedded in.

The next major step in our study will be to explore in interviews and informal discussions with domain experts from the field of invasion biology as well as through ethnographic observations in an ongoing study, the relationship between the topic structures constructed by the chosen algorithmic approaches with the lived experience of researchers in the field of invasion biology. Specifically, we plan to pursue the following avenues:

1. How do individual research group leaders' research trails (Gläser & Laudel 2015) relate to the topical structures of the Leiden[16] and projection solutions? Do research topics that can be delineated within those trails align with or transcend the field topics we have constructed algorithmically?

2. Existing, theoretical work on empirical evidence in the field of invasion biology identifies theoretical work on key hypothesis of the field as well as empirical studies that support or challenge those hypotheses[8]. This offers the opportunity to relate the topical structures we have generated to relevant topical perspectives generated from within the field and discuss these relations with domain experts.

---

[8] https://hi-knowledge.org/

**Figure 2: Topic affinity networks[9] of a) Leiden[16] solution and b) projection solution.**

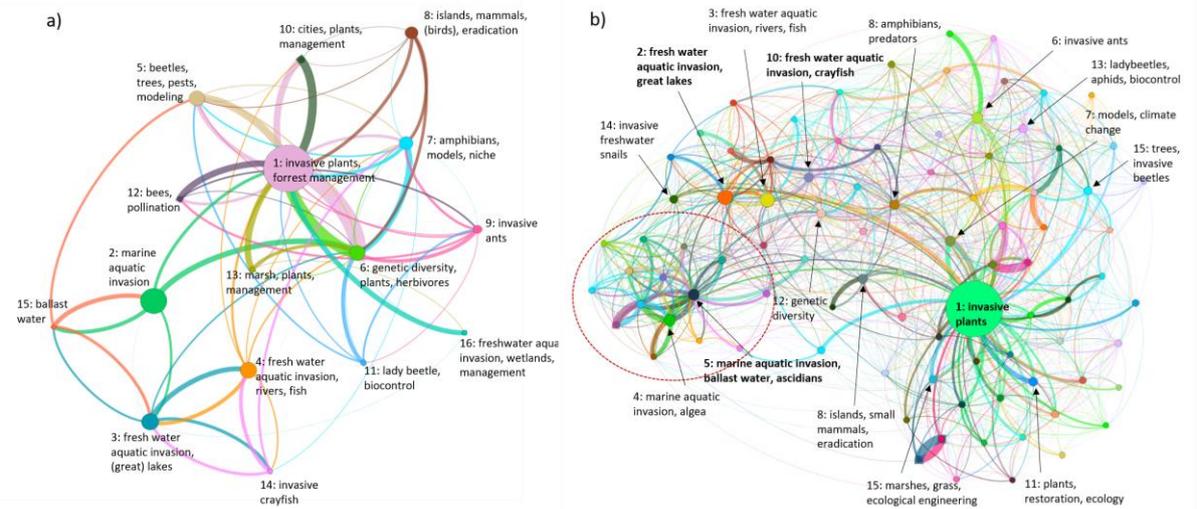

**Figure 3: Alluvial showing flow between clusters in Leiden[16] and Projection 91.**

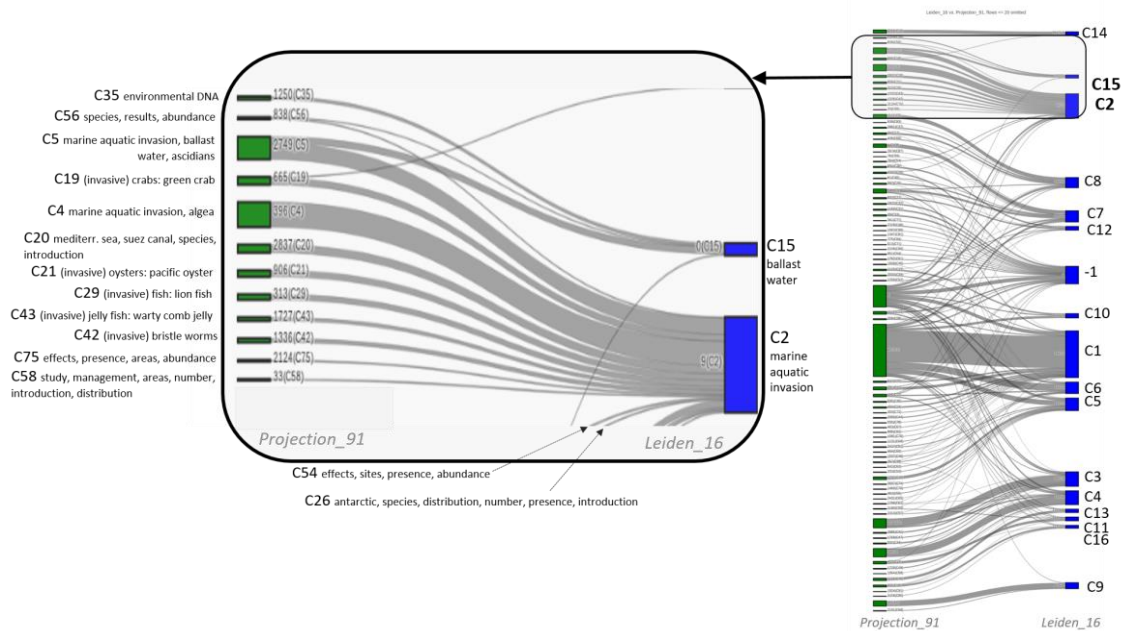

**Conclusions**

We report first results on a comparison of an internal and an external mapping of topical structures in the research specialty of invasion biology. Both maps exhibit some common features, like the importance of work on invasive plants, and the relevance of concepts of habitat and organism for distinguishing topics. The next step in the study will be dedicated to relating the algorithmically identified topic structures to topical concepts emerging from social and theoretical processes within the research specialty.

---

[9] Node size reflects number of publications (viz. gephi), links reflect disproportionately strong affinity. Link curvature indicates link direction (clockwise).

**Acknowledgments** This research of M. Held was supported by the German Federal Ministry of Education and Research (Grant 01PU17003).

# Data Appendix

**Table 1. Descriptions of the topical content of the clusters in the Leiden solutions.**

| Cluster (Size) | Size | Topic | Ecosystem Type |
|---|---|---|---|
| Leiden$_6$ C1 | 8769 | (invasive) plants; (restoration) ecology, vegetation science | terrestrial |
| Leiden$_6$ C2 | 4542 | freshwater, great lakes; fish, mussel, crustacean; hydrobiology, fisheries | aquatic |
| Leiden$_6$ C3 | 3535 | marine; crustaceans, mussels, algea; ballast water; marine biology | aquatic |
| Leiden$_6$ C4 | 2949 | islands; mammals, birds, amphibians; conservation; wildlife research | hybrid |
| Leiden$_6$ C5 | 2104 | trees, beetles; pests, modeling; economics, entomology | terrestrial |
| Leiden$_6$ C6 | 1308 | ants, bees; pollination; entomology, insects sociaux | terrestrial |
| Leiden$_{16}$ C1 | 5822 | (invasive) plants; (restoration) ecology, vegetation science, forest management | terrestrial |
| Leiden$_{16}$ C2 | 3051 | marine, coast; crustaceans, mussels, algea; aquatic invasion, aquaculture; marine biology | aquatic |
| Leiden$_{16}$ C3 | 1897 | freshwater, great lakes; round goby, zebra mussel, zooplankton; aquatic invasion; hydrobiology | aquatic |
| Leiden$_{16}$ C4 | 1851 | freshwater, river; fish; assemblage; fish biology | aquatic |
| Leiden$_{16}$ C5 | 1719 | trees, beetles; pests, modeling; economics, entomology | terrestrial |
| Leiden$_{16}$ C6 | 1680 | plants, herbivores; genetic diversity, evolution, invasive populations; molecular ecology | terrestrial |
| Leiden$_{16}$ C7 | 1486 | amphibians; models, niche; herpetology, behavioral ecology | hybrid |
| Leiden$_{16}$ C8 | 1369 | islands; mammals (birds); conservation; wildlife research | terrestrial |
| Leiden$_{16}$ C9 | 863 | (invasive) ants; myrmecology, sociobiology, environmental/ecological entomology | terrestrial |
| Leiden$_{16}$ C10 | 657 | cities; plants; management; urban planning | terrestrial |
| Leiden$_{16}$ C11 | 537 | lady beetle, aphids; biocontrol, (environmental, ecological) entomology | terrestrial |
| Leiden$_{16}$ C12 | 523 | bees, plants; pollination; apidology, ecology | terrestrial |
| Leiden$_{16}$ C13 | 505 | wetlands; grass, plant; ecology, management, aquatic botany | hybrid |
| Leiden$_{16}$ C14 | 485 | freshwater; (invasive) crayfish; management of aquatic ecosystems, freshwater biology | aquatic |
| Leiden$_{16}$ C15 | 436 | ballast water; environmental science | aquatic |
| Leiden$_{16}$ C16 | 419 | freshwater, wetlands; aquatic plant; botany, management | aquatic |

**Table 2. Descriptions of the topical content of the 16 largest projection clusters and information on the microfields the projection clusters are embedded in.**

| Cluster (Size) | Topic | Eco-system Type | Microfield (size) share | Microfield RAs[10] (# publications) |
|---|---|---|---|---|
| Projection C1 (6308) | (invasive) plants; invasive plant science, ecology, management | terrestrial | m402 (11978) **52.7%** | Ecology (6071) Plant Sciences (2722) Biodiversity Conservation (2128) |
| Projection C2 (1136) | freshwater, great lakes; round goby, zebra mussle, zooplankton; aquatic invasion; hydrobiology | aquatic | m1174 (6689) **17%** | Marine & Freshwater Biology (3336) Fisheries (2325) Limnology (1362) |
| Projection C3 (1042) | freshwater; fish; ecology, biology, fisheries | aquatic | m34 (21481) 4.9%) | Marine & Freshwater Biology (7989) Ecology (5563) Environmental Sciences (4771) |
| Projection C4 (773) | marine; algae; marine biology and ecology, phycology | aquatic | m396 (12027) 6.4% | Marine & Freshwater Biology (7536) Plant Sciences (2847) Ecology (2804) |
| Projection C5 (708) | marine; ascidians; ballast water, aquatic invasion; ecology, biology, policy | aquatic | m2749 (6689) **34%** | Marine & Freshwater Biology (780) Ecology (539) Environmental Sciences (295) |
| Projection C6 (665) | (invasive) ants; sociobiology, environmental and ecological entomology | terrestrial | m351 (12582) 5.3% | Entomology (4944) Ecology (3287) Zoology (2205) |
| Projection C7 (589) | climate change; potential distribution; biogeography, biodiversity, conservation | N.A. | m149 (16119) 3.7% | Ecology (9437) Biodiversity Conservation (4348) Environmental Sciences (3473) |
| Projection C8 (500) | amphibians; invasive species, predators; herpetology, conservation | hybrid | m447 (11504) 4.3% | Zoology (4834) Ecology (3690) Environmental Sciences (1335) |
| Projection C9 (481) | islands; (small) mammals; eradication; ecology, wildlife research | terrestrial | m1112 (6973) 6.9% | Ecology (3520) Zoology (2757) Biodiversity Conservation (1060) |
| Projection C10 (417) | freshwater; crayfish; aquatic ecosystems management, freshwater biology | aquatic | m2568 (2481) **16.8%** | Marine & Freshwater Biology (849) Fisheries (510) Ecology (442) |
| Projection C11 (394) | plant; communities, diversity, richness; vegetation science, restoration ecology | terrestrial | m391 (12080) 3.3% | Ecology (7597) Plant Sciences (4338) Forestry (1722) |
| Projection C12 (364) | genetic diversity; conservation genetics, molecular ecology | N.A. | m126 (16786) 2.2% | Ecology (5223) Genetics & Heredity (4689) Evolutionary Biology (4232) |
| Projection C13 (350) | ladybeetles, aphids; biocontrol; (economic) entomology | terrestrial | m427 (11710) 3.0% | Entomology (7034) Ecology (1744) Biotechnology & Applied Microbiology (1034) |

---

[10]Top three Web of Science research areas in the microfields. Note that some journals are assigned to more than one RA.

| Projection C14 (349) | (invasive) freshwater snails; molluscan research | aquatic | m1110 (6987) 5.0% | Zoology (3247) Marine & Freshwater Biology (2628) Ecology (1118) |
| --- | --- | --- | --- | --- |
| Projection C15 (345) | trees; (invasive) beetles; economic entomology | terrestrial | m1737 (4695) 7.3% | Entomology (1750) Forestry (914) Ecology (581) |
| Projection C16 (325) | marshes, wetlands; grass; ecological engineering | hybrid | m1108 (6988) 4.7% | Environmental Sciences (2402) Ecology (1916) Marine & Freshwater Biology (1751) |